\documentclass[conference]{IEEEtran}
\IEEEoverridecommandlockouts
\usepackage{cite}
\usepackage[linesnumbered,ruled,vlined]{algorithm2e}
\usepackage{amsmath,amssymb,amsfonts}
\usepackage{algorithmic}
\usepackage{graphicx}
\usepackage{textcomp}
\usepackage{xcolor}
\usepackage{tikz}
\usepackage{threeparttable}
\usepackage{booktabs} 
\usepackage{subcaption} 
\usepackage{array}
\def\BibTeX{{\rm B\kern-.05em{\sc i\kern-.025em b}\kern-.08em
    T\kern-.1667em\lower.7ex\hbox{E}\kern-.125emX}}

\IEEEpubid{\makebox[\columnwidth]{979-8-3315-2710-5/25/\$31.00 \copyright~2025 IEEE \hfill \hspace{\columnsep} \makebox[\columnwidth]{}}}

\begin{document}


\title{\textcolor{black}{DIRC-RAG: Accelerating Edge RAG with Robust High-Density and High-Loading-Bandwidth\\Digital In-ReRAM Computation}\vspace{-2mm}}

 \author{\IEEEauthorblockN{Kunming Shao\textsuperscript{1,*,†}, Zhipeng Liao\textsuperscript{2,*}, Jiangnan Yu\textsuperscript{1,*}, Liang Zhao\textsuperscript{3}, Qiwei Li\textsuperscript{4}, Xijie Huang\textsuperscript{1}, Jingyu He\textsuperscript{1}, \\Fengshi Tian\textsuperscript{1}, Yi Zou\textsuperscript{3}, Xiaomeng Wang\textsuperscript{1,†}, Tim Kwang-Ting Cheng\textsuperscript{1}, Chi-Ying Tsui\textsuperscript{1}}



 \thanks{This research was supported by ACCESS - AI Chip Center for Emerging Smart Systems, sponsored by InnoHK funding, Hong Kong SAR, partially supported by SCUT Research Fund No. K3200890, as well as partially by Guangzhou GJYC Fund No. 2024D01J0010.}

 \IEEEauthorblockA{
 \textsuperscript{1}The Hong Kong University of Science and Technology
 \textsuperscript{2}University of Southampton\\
 \textsuperscript{3}South China University of Technology
 \textsuperscript{4}Wuhan University\\
 *Equally Contributed Authors. †Correspondence: kshaoaa@connect.ust.hk and xwangee@connect.ust.hk.
 }


 
\vspace{-7mm}
}

\maketitle

\begin{abstract}

Retrieval-Augmented Generation (RAG) enhances large language models (LLMs) by integrating external knowledge retrieval but faces challenges on edge devices due to high storage, energy, and latency demands. Computing-in-Memory (CIM) offers a promising solution by storing document embeddings in CIM macros and enabling in-situ parallel retrievals but is constrained by either low memory density or limited computational accuracy. To address these challenges, we present DIRC-RAG, a novel edge RAG acceleration architecture leveraging Digital In-ReRAM Computation (DIRC). DIRC integrates a high-density multi-level ReRAM subarray with an SRAM cell, utilizing SRAM and differential sensing for robust ReRAM readout and digital multiply-accumulate (MAC) operations. By storing all document embeddings within the CIM macro, DIRC achieves ultra-low-power, single-cycle data loading, substantially reducing both energy consumption and latency compared to off-chip DRAM. A query-stationary (QS) dataflow is supported for RAG tasks, minimizing on-chip data movement and reducing SRAM buffer requirements. We introduce error optimization for the DIRC ReRAM-SRAM cell by extracting the bit-wise spatial error distribution of the ReRAM subarray and applying targeted bit-wise data remapping. An error detection circuit is also implemented to enhance readout resilience against device- and circuit-level variations.

Simulation results demonstrate that DIRC-RAG under TSMC 40nm process achieves an on-chip non-volatile memory density of 5.18Mb/mm\textsuperscript{2} and a throughput of 131 TOPS. It delivers a 4MB retrieval latency of 5.6$\mu$s/query and an energy consumption of 0.956$\mu$J/query, while maintaining the retrieval precision.


\end{abstract}


\section{Introduction}


Large Language Models (LLMs), relying on extensive parameters and diverse training data, have demonstrated exceptional capabilities in language understanding and generation. However, for certain user-defined private data, such as medical records or personal information, privacy concerns prevent these data from being transmitted to the cloud for processing. At the same time, performing costly retraining or fine-tuning locally is not feasible. A secure and efficient solution is to integrate personalized knowledge documents into LLM inference using edge Retrieval-Augmented Generation (RAG) technique. Edge RAG ensures data privacy while significantly improving the model's performance across various domains without the need for expensive fine-tuning\cite{nips2020retrieval, arxiv2024retrieval, ICLR2024corrective, ren2024retrieval}.

\begin{figure}[t]
    \centering
    \includegraphics[width=\columnwidth]{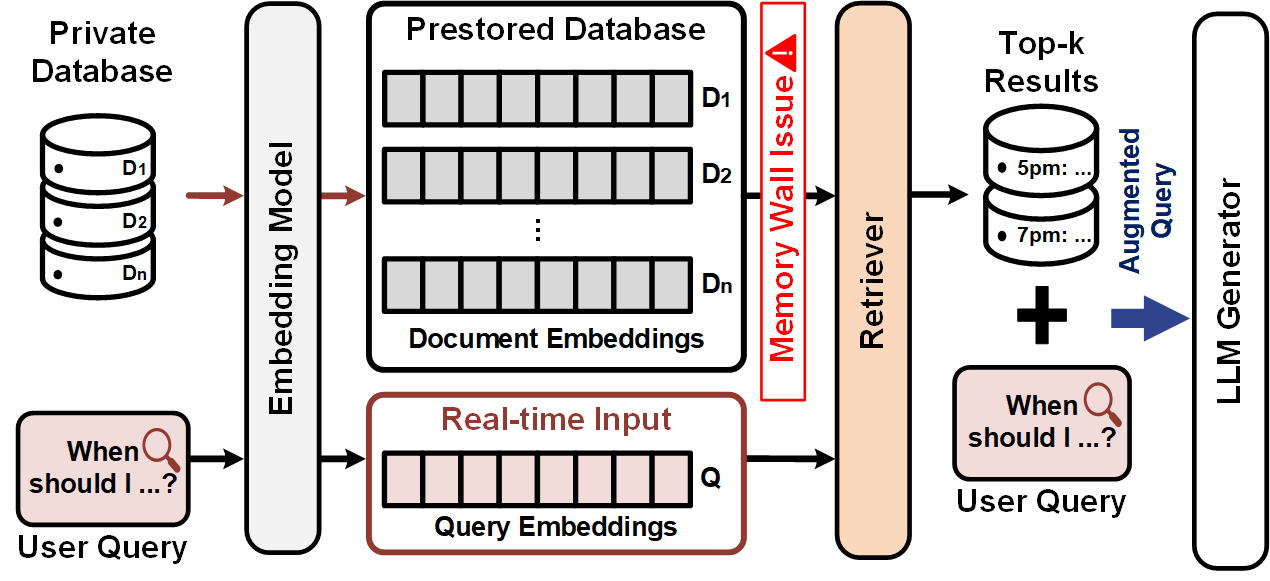}
    \caption{\small{Demonstration of the workflow of RAG, where the document embedding storage and loading for retrieval are the bottlenecks.}}
    \vspace{-4mm}
    \label{fig1}
    \end{figure}

As illustrated in Figure \ref{fig1}, the private database is first converted into document embeddings using an embedding model and stored locally. When a user submits a real-time query, it is transformed into a query embedding, which is then compared with the pre-stored document embeddings using cosine similarity or Maximum Inner Product Search (MIPS). The most relevant document chunks are retrieved and, along with the original query, input into the LLM for augmented inference and generation. However, the storage and loading of large-scale document embeddings and retrieval computation with query embeddings pose significant challenges to hardware systems in terms of storage capacity, energy consumption, and latency\cite{shen2024towards, qin2024robust, quinn2024accelerating, issccrag}.

Computing-in-Memory (CIM) architectures \cite{romcim1,reramcim1,reramcim2,sramcim1,sramcim4,edram1,edram2} have been proposed as a promising solution to address the memory wall issue by caching weights on CIM macros and adopting a weight-stationary dataflow. CIM integrates computation logics within memory arrays, enabling in-situ parallel computation and significantly reducing data movement overhead. This architecture is particularly well-suited for workloads dominated by multiply-accumulate (MAC) operation, such as those found in neural networks and other AI applications. By leveraging CIM, large-scale document embeddings can be stored in CIM macro, query embedding is taken as input, and the retrieval MAC operation is processed directly within memory array, minimizing the need for frequent data movement between off-chip memory and on-chip processing units. This not only reduces latency but also eliminates the energy overhead associated with data loading, making CIM an attractive option for accelerating RAG workloads on edge devices. 


However, current CIM architectures remain constrained by their limited storage density and update bandwidth for memory-intensive RAG retrieval operations. To address the above challenges, we propose DIRC-RAG, a novel edge RAG acceleration architecture based on Digital In-ReRAM Computation (DIRC). DIRC combines the high-density storage capabilities of ReRAM with the robust and efficient computation capabilities of SRAM, providing a high-density, low-latency solution optimized for RAG workloads\cite{dac2022res,iccad2023res,iccad24rescim}. Compared to existing CIM architectures, DIRC-RAG offers several key advantages tailored to RAG tasks:
\begin{itemize}
    \item \textbf{ReRAM and SRAM Coupled Memory:} By coupling high-density ReRAM subarrays with robust SRAM cells using a differential sensing circuit, DIRC achieves significantly higher memory density and reliability, addressing the storage and loading demands of large-scale document embeddings.
    \item \textbf{Query Stationary Dataflow:} Leveraging the ultra-low-power and one-cycle-latency data-loading capability of DIRC macro, DIRC-RAG could support the query-stationary dataflow for retrieval, minimizing on-chip data movement and minimizing SRAM buffer requirements.
    \item \textbf{Digital Multiply-Accumulation (MAC):} DIRC integrates 128x128 ReRAM-SRAM coupled DIRC cells with bitwise multipliers, 128-input carry-save-adders and accumulators, enabling high-throughput and efficient digital MAC operations for various INT precisions.
\end{itemize}

Additionally, DIRC introduces error-aware optimization techniques to address the variability of both ReRAM devices and CMOS circuits. Specifically:
\begin{itemize}
    \item \textbf{Error-Aware Bitwise Mapping:} Through detailed post-layout Monte-Carlo simulation, a ReRAM subarray bitwise spatial read-out error map was generated based on ReRAM deviation and MOS mismatch. This map guides the bitwise mapping of the most significant bits (MSBs) in the most robust ReRAM positions, while the less stable positions are for the least significant bits (LSBs).
    \item \textbf{Error Detection after Sensing:} An error detection circuit is embedded to check the read-out error of the ReRAM-SRAM coupled cell, and if an error is detected, the data is re-sensed to ensure accuracy.
\end{itemize}

The remainder of this paper is structured as follows: Section II presents the Preliminaries on RAG, CIM and non-volatile CIM; Section III demonstrates the overall architecture of DIRC-RAG; Section IV showcases the experimental results of DIRC-RAG. Finally, Section V concludes the paper.

\section{Preliminaries}

\subsection{Retrieval-Augmented Generation}



RAG is a technique that integrates external knowledge into pre-trained LLMs in a non-parametric manner, significantly enhancing their performance in domain-specific tasks.
As illustrated in Figure \ref{fig1}, RAG employs embedding models to convert textual data into high-dimensional vector representations (embeddings), which are then passed to a retriever for vector-based retrieval. Common retrieval methods include Cosine Similarity and Maximum Inner Product Search (MIPS). The choice of method depends on the output form of the embedding model: if the embeddings are normalized, Cosine Similarity is preferred; if the embeddings are unnormalized, MIPS can be used and computed by vector inner products.
\vspace{-2mm}
\[
Q = \begin{bmatrix} q_1, q_2, q_3, \dots, q_{n} \end{bmatrix}, \quad D = \begin{bmatrix} d_1, d_2, d_3, \dots, d_{n} \end{bmatrix}
\]
\vspace{-4mm}
\[
\text{Inner\_Product}(Q, D) = \sum_{i=1}^{n} q_i d_i
\]
\vspace{-1mm}
\[
\text{Cosine\_Similarity}(Q, D) = \frac{\sum_{i=1}^{n} q_i d_i}{\sqrt{\sum_{i=1}^{n} q_i^2} \cdot \sqrt{\sum_{i=1}^{n} d_i^2}}
\]
Both Cosine Similarity and MIPS rely on vector dot product computations, often referred to as MAC operations. To accelerate this computation pattern, CIM architectures can be employed, offering significant performance improvements for such tasks.

\subsection{Mainstream Computing-in-Memory Technologies}
\begin{figure}[t]
\centering
\includegraphics[width=\columnwidth]{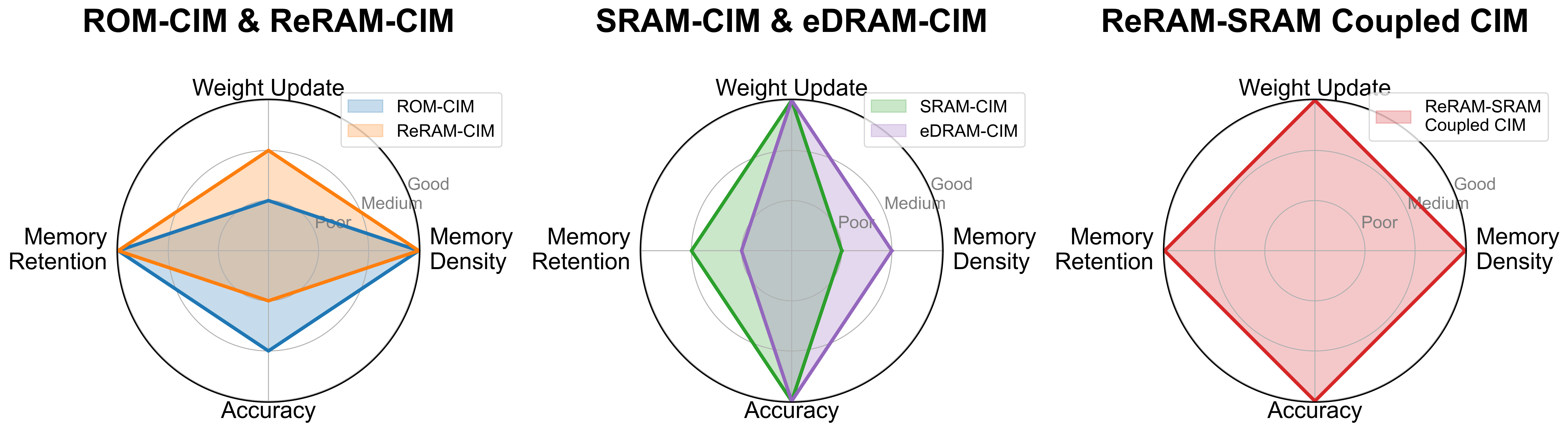}
\caption{\small{Comparison between mainstream CIM memories.}}
\vspace{-6mm}
\label{fig2}
\end{figure}

Figure \ref{fig2} introduces four mainstream memory technologies for Computing-In-Memory (CIM):

\begin{itemize}
    \item \textbf{ROM-CIM:} Utilizes fixed ROM with parameters set during fabrication, unsuitable for dynamic updates\cite{romcim1}.
    \item \textbf{ReRAM-CIM:} Based on non-volatile resistive RAM, offering high density and rewritability. However, analog computation in mainstream designs suffers from deviations (e.g., resistance drift), reducing accuracy \cite{reramcim1, reramcim2}.
    \item \textbf{SRAM-CIM:} Leverages static RAM with high computational accuracy but limited storage density due to its complex 6T or larger cell structure \cite{sramcim1, sramcim4}.
    \item \textbf{eDRAM-CIM:} Typically based on 3T1C embedded DRAM, combining high accuracy and density. However, periodic refresh increases power consumption and latency, limiting efficiency in edge devices \cite{edram1, edram2}.
\end{itemize}

\begin{figure*}[!t]
    \centering
    \includegraphics[width=\textwidth]{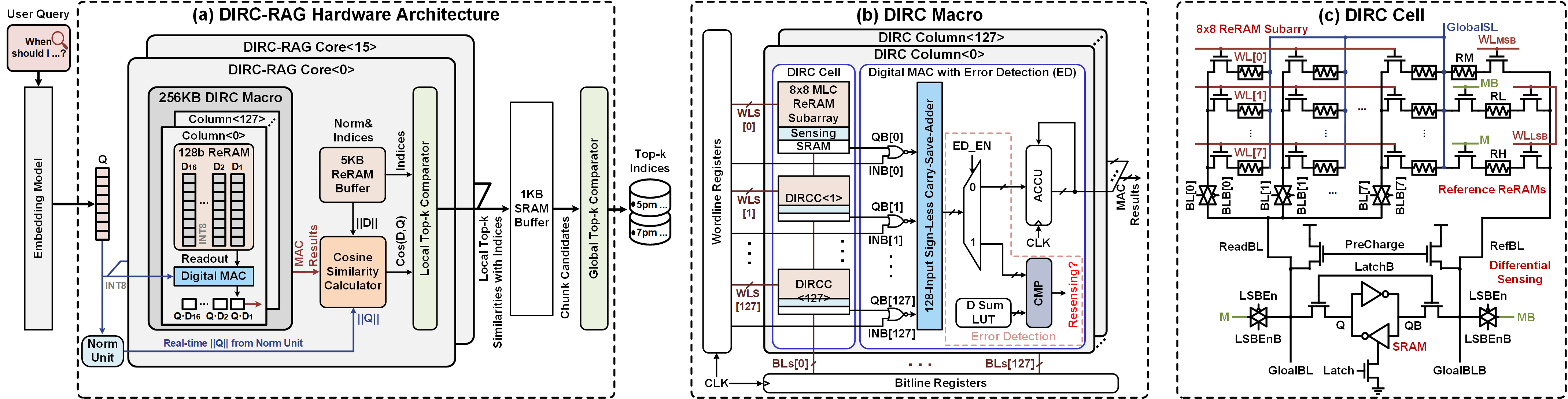}
    \caption{\small{ (a) DIRC-RAG architecture for document retrieval with query-stationary dataflow. (b) DIRC macro with 128x128 DIRC cells, digital MAC and error detection circuit.  (c) DIRC cell with 8x8 MLC ReRAM subarray, differential sensing circuit and 1bit SRAM cell.}}
    \vspace{-4mm}
    \label{fig3}
    \end{figure*}

\subsection{ReRAM-SRAM Coupled CIM}
To address the limitations of mainstream CIM architectures in storage density and computational accuracy, a novel ReRAM-SRAM Coupled CIM paradigm has been proposed, balancing high storage density and computational precision by coupling ReRAM with SRAM-CIM.
\cite{dac2022res} firstly integrated ReRAM into SRAM-CIM by embedding multiple SLC ReRAMs into SRAM cells. Weight data stored in ReRAM is loaded into SRAM via current-based sensing, significantly improving storage density.
\cite{iccad2023res} introduced a two-SRAM-cell, triple-level ReRAM-SRAM Coupled ternary CIM architecture, improving both storage density and data loading accuracy.
\cite{iccad24rescim} supported four-level MLC ReRAM with a single SRAM cell, boosting storage density and efficiency, while proposing a robust differential sensing scheme for tackling data loading errors.

\section{DIRC-RAG Framework}

\subsection{Overall Architecture}
Figure \ref{fig3} illustrates the architectural details of the DIRC-RAG, including the DIRC-RAG architecture, the DIRC macro circuit, and the transistor-level design of the DIRC cell.

As shown in Figure 3(a), the DIRC Architecture consists of sixteen DIRC-RAG Cores, where each core independently stores and processes different document embeddings. The architecture also includes a norm unit for real-time computation of the query embedding vector norm to support cosine similarity calculations, an SRAM buffer to store the similarity scores and indices of the local top-k results produced by each core, and a Global Top-k Comparator that compares the local results from sixteen cores and outputs the indices corresponding to the final top-k results.
Each core internally comprises a DIRC macro that performs document embedding storage and MAC operations, a ReRAM buffer for storing the norms and indices of document embeddings, a calculator for computing cosine similarity that can be bypassed when MIPS, and a local top-k comparator that executes top-k selection within the core.

The sixteen DIRC-RAG Cores operate in parallel and independently. When a query embedding is input, it is broadcast to all sixteen cores. Each core then computes the cosine similarity or inner product between the query embedding and all document embeddings stored within the core. The results are compared locally within each core to select the local top-k results, which are subsequently stored in the SRAM buffer. Since the local top-k selection eliminates the majority of candidates, the SRAM buffer's storage requirements are minimal. Finally, the Global Top-k Comparator selects the most similar document embeddings from the SRAM buffer and outputs their corresponding indices.

As shown in Figure 3(b), a DIRC macro consists of 128 DIRC columns and peripheral input registers. Each DIRC column is composed of the following components: 128 vertically aligned DIRC cells with NVM storage; 128 NOR gates functioning as bit multipliers; a high-speed and efficient 128-input sign-less carry-save adder\cite{csa0,csa1,csa2}; and an accumulator circuit that performs cycle-by-cycle accumulation for adder output.
Additionally, to address potential ReRAM readout errors, each DIRC column is equipped with an optional Error Detection Circuit designed to detect errors in the SRAM cached data. During the error detection cycle, the peripheral input registers input 128 logical `1`s, causing the adder to output the sum of all D data in the column, denoted as $\Sigma D$. This result is then compared to the pre-stored value in the D Sum Look-up Table. If the comparison results differ, an error is detected, and the DIRC cell will re-sense the data to ensure computational accuracy.
The high density and large capacity of the DIRC macro enable a storage capacity of 2Mb and a storage density of up to 5.88Mb/mm\textsuperscript{2}. The error detection circuit further ensures the accuracy of computations, making the design both reliable and efficient. Moreover, the SRAM can be written from outside data, which means that if the ReRAM is not large enough for stoarge, the computational part of DIRC macro can be used as a general SRAM-CIM macro.

Refer to \cite{iccad24rescim}, we build a DIRC cell as shown in Figure 3(c). The DIRC cell consists of the following components: an 8x8 MLC ReRAM subarray in the top-left corner for data storage; three reference ReRAMs in the top-right corner to store intermediate reference values that distinguish four ReRAM resistance levels; a differential sensing circuit in the middle, which performs readout through Latch and Precharge; and a 1-bit SRAM cell at the bottom to store the readout result.
During the differential sensing process, the Latch signal is first disabled, disconnecting the feedback loop of the SRAM. The Precharge circuit charges the Q and QB nodes of the SRAM to $V_{DD}/2$. Subsequently, the Latch circuit is enabled, restoring the feedback loop of the SRAM, which discharges the read bitline (ReadBL) and reference bitline (RefBL). The discharging rate depends on the relative loads of the two bitlines.

For the MLC ReRAM to be read from the 8x8 subarray, its MSB is first sensed to determine whether its resistance value resides in the lower two levels or the higher two levels. Specifically, the GlobalSL is set to 0, and the corresponding WL and BL in the subarray are selected, and the WL\textsubscript{MSB} in the Reference is also selected, creating two relative grounded loads on both ReadBL and RefBL. When the voltage at the Q and QB nodes is $V_{DD}/2$, the side with lower load on bitline discharges faster to 0, while the other side with higher load is charged to $V_{DD}$ by the feedback loop. If the ReRAM resistance is smaller than $R_{M}$, the Q node of the SRAM discharges to 0; otherwise, the Q node charges to $V_{DD}$.
For LSB sensing, the LSBEn signal is enabled, and based on the MSB sensing result of the previous cycle, the M and MB signals select either $R_{L}$ or $R_{H}$ as the reference. The same differential sensing process is repeated and the LSB is read. Differential sensing effectively reduces the impact of ReRAM variations on computational accuracy and makes it feasible to embed logic circuits in high-density ReRAM arrays\cite{natureelec2018resistive}.

\subsection{Query-Stationary Dataflow}

\begin{figure}[t]
    \centering
    \includegraphics[width=0.95\columnwidth]{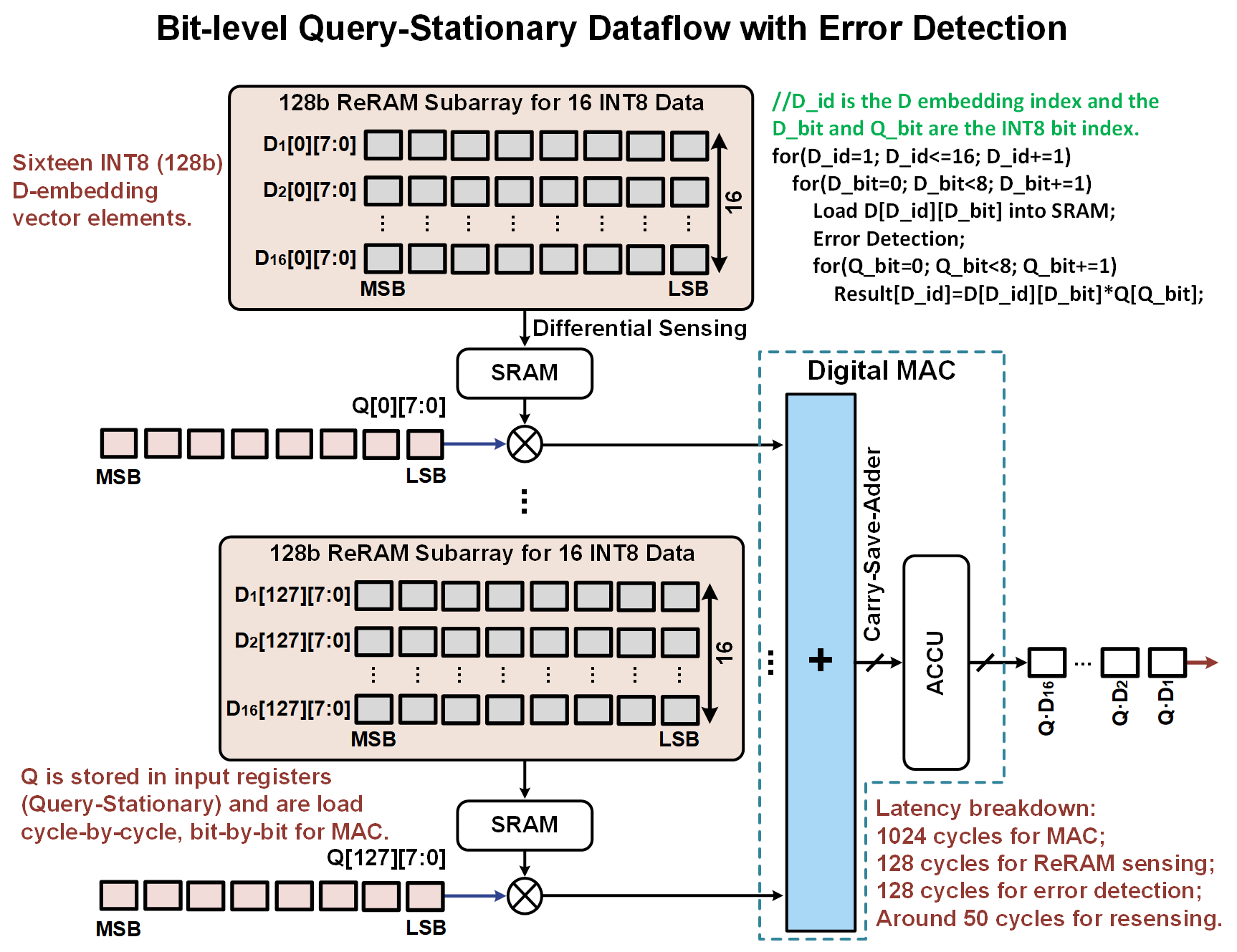}
    \caption{\small{Bit-level query-stationary dataflow in a DIRC column, which stores sixteen document embeddings and enables error detection after ReRAM sensing.}}
    \vspace{-2mm}
    \label{fig4}
    \end{figure}
Taking SRAM-CIM as an example, mainstream CIM technologies typically adopt weight stationary (WS) dataflow to enhance the reuse of weight data in DNN/Transformer algorithms. However, for retrieval tasks, where a single query embedding is matched against thousands or millions of document embeddings for MAC computations, two major dataflow challenges arise:

\begin{itemize}
    \item \textbf{Storage limitation with WS: } Storing document embeddings in the SRAM of CIM macros using WS dataflow is constrained by limited SRAM density. This prevents the CIM macro from accommodating all document embeddings, requiring tens to hundreds of cycles for row-by-row SRAM updates after a few MAC computation cycles, which significantly increases latency and energy consumption.
    
    \item \textbf{Low utilization with IS: } Using an input stationary (IS) dataflow, as proposed in \cite{hpca2023inca, iccd2024hpa}, results in low CIM macro utilization due to the small number of query embeddings (usually one). Each retrieval also incurs high memory access overhead from on-chip/off-chip buffers to the CIM macro, further increasing computational costs.
\end{itemize}

To address these challenges, DIRC adopts a fitting query-stationary dataflow, effectively overcoming the dataflow issues faced by CIM in retrieval tasks. First, DIRC utilizes Non-volatile MLC ReRAM, which offers significantly higher memory density than SRAM-CIM, enabling long-term storage of a large number of document embeddings that require infrequent updates. Second, it employs In-ReRAM Computation, achieving ultra-low power consumption and enabling data to be loaded from ReRAM into all SRAMs in the array within a single cycle. Finally, DIRC stores the limited query embeddings in sufficient input registers, eliminating the need for on-chip buffer access until the retrieval process is complete.

If the document embedding dimension exceeds 128, it can be folded and mapped within the same DIRC column. For instance, a single DIRC column can store sixteen INT8 embeddings with a dimension of 128 or two INT8 embeddings with a dimension of 1024. Additionally, DIRC can store twice as many INT4 quantized embeddings as INT8 quantized embeddings in a single column.

As shown in Figure 4, each DIRC column can hold sixteen INT8 quantized document embeddings with a dimension of 128. The SRAM reads 1-bit of the INT8 document embedding at a time and performs cycle-by-cycle MAC computations with the 8-bit serial input query embedding. Partial sums generated in each cycle are accumulated by an accumulator. Since all DIRC macros operate in parallel, the vector dot product between the query embedding and all 4MB document embeddings can be computed in approximately 1300 cycles, or about $5.2\mu$s at a 250MHz frequency. By adopting the query-stationary dataflow, DIRC-RAG achieves efficient and low-latency retrieval.
\begin{figure}[t]
    \centering
    \includegraphics[width=\columnwidth]{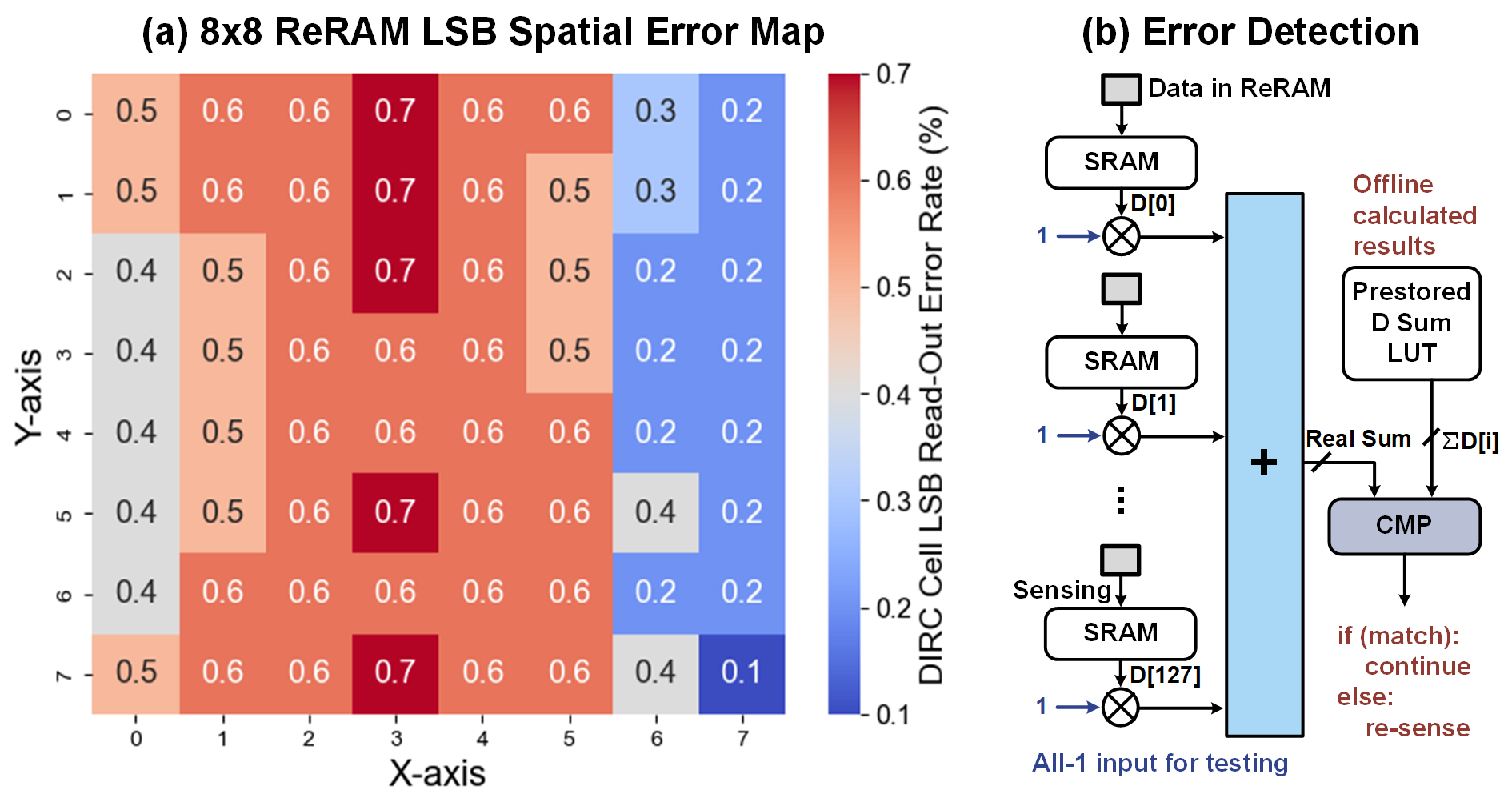}
    \vspace{-6mm}
    \caption{\small{(a) The LSB spatial error map of the 8x8 ReRAM subarray from post-layout Monte Carlo simulation. (b) The error detection circuit for checking the sensing error of the DIRC cell.}}
    \vspace{-4mm}
    \label{fig5}
    \end{figure}

\subsection{Error-Aware Optimization Techniques}
\begin{figure}[t]
    \centering
    \includegraphics[width=\columnwidth]{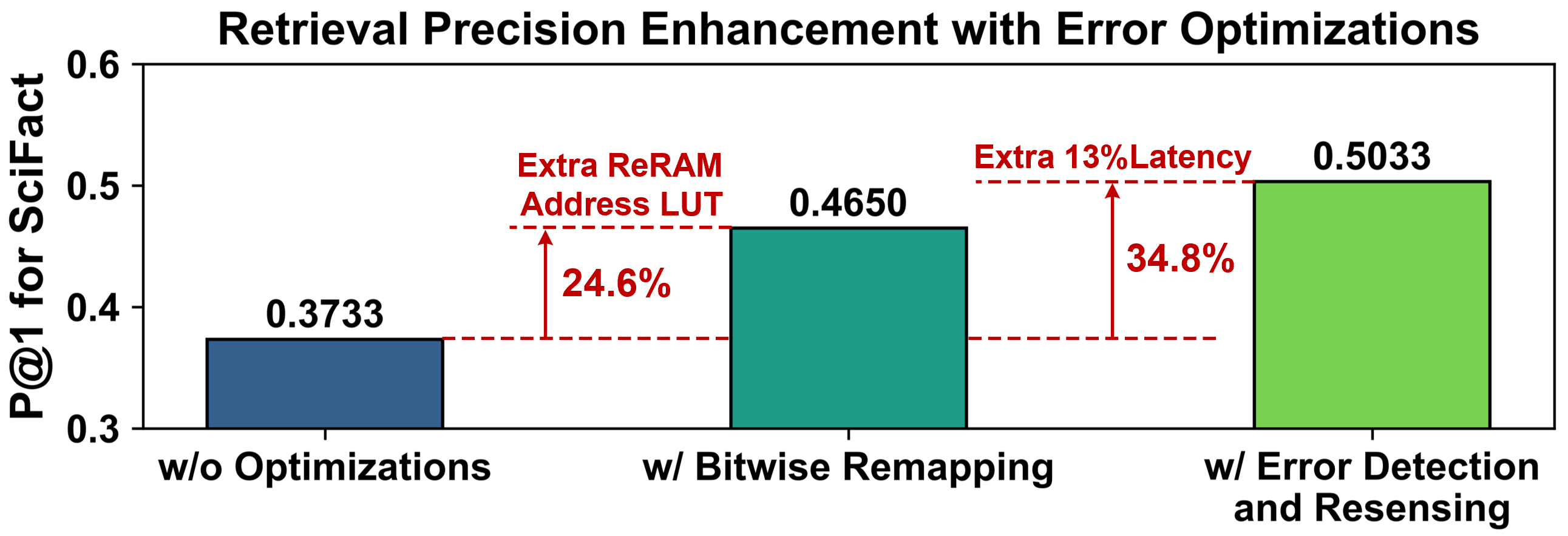}
    \vspace{-6mm}
    \caption{\small{The effectiveness of the error-aware optimization techniques for increasing the retrieval precision.}}
    \vspace{-6mm}
    \label{fig6}
    \end{figure}
Even though the differential sensing scheme adopted by the DIRC cell effectively reduces the impact of ReRAM variation on computational accuracy \cite{iccad24rescim}, potential errors in high-density ReRAM still arise due to outlier ReRAM deviations and MOS process mismatches. These errors manifest as unpredictable bit flip errors during readout operations.

To address these bit flip errors, we conducted a 1000-point post-layout Monte Carlo simulation for the DIRC cell. The simulation introduced ReRAM deviations ($\sigma=0.1$) and MOS mismatches under 0.8V and 250MHz, considering the actual spatial distribution of ReRAM, voltage supply, and routing constraints. The resulting 8x8 subarray LSB spatial error map is shown in Figure 5(a). The MSB of MLC ReRAM demonstrated 100\% reliability due to its large signal margin, so we focused on the LSB error distribution. In Figure 5(a), two VSS power rails were placed on the left and right sides of the subarray for NMOS transistors, while the sensing circuit and SRAM were on the right side. ReRAM cells closer to the VSS rail exhibited smaller read errors, while those farther from the readout circuit exhibited larger errors, resulting in the observed spatial error pattern.

Given the significant differences in error distribution, we adopted a bit-wise remapping strategy. INT8 data is divided into two groups: bits 0-3 and bits 4-7 (where bit 7 is the sign bit). Since MSBs showed 100\% reliability, bits 4-7 were mapped to MSBs, while errors in bits 0-3, located on LSBs, were considered. Using the LSB spatial error map, bit 3 of the sixteen INT8 data was mapped to locations with the smallest error rates, while bit 0 was mapped to locations with the largest error rates. This strategy minimizes the impact of errors on computational accuracy. Figure 6 demonstrates the 24.6\% precision improvement achieved by the bitwise remapping.

Since bit flip errors are unpredictable and cannot be compensated for or retrained as proposed in \cite{qin2024robust}, we designed an error detection circuit, shown in Figure 5(b), to identify sensing errors in the DIRC cell. Assuming correct data is written into ReRAM, the circuit detects bit flip errors caused by sensing issues. To support this, DIRC-RAG computes the bitwise sum of document embeddings offline and stores the results in a D Sum Look-Up Table (LUT) based on the ReRAM buffer. During runtime, LUT values are compared with the adder output. If differences occur, an error is detected, and the DIRC cell re-senses the data to ensure accuracy. This method could effectively mitigate errors caused by transient interference during sensing, as shown in Figure 6.

\section{Experiments and Validation}
\subsection{Experiment Setup}

For the hardware experiments, we implemented a DIRC macro using the TSMC 40nm process for post-layout simulation. The digital MAC and error detection circuits were synthesized with Synopsys Design Compiler, while Cadence Innovus was used for automatic placement and routing. The DIRC cell was designed and verified in Cadence Virtuoso, where the top-level layout and post-layout simulations were also conducted. The ReRAM model is adopted from \cite{reram2017nc} and placed in the VIA1 layer. The SRAM buffer is generated using the TSMC40nm SRAM compiler, while the remaining digital circuits are synthesized with Synopsys Design Compiler and simulated in Synopsys VCS.

\begin{table}[t]
    \centering
    \caption{DIRC-RAG SPEC}
    \vspace{-2mm}
    \resizebox{0.7\columnwidth}{!}{ 
    \begin{tabular}{|l|c|}
        \hline
        \textbf{Process} & TSMC40nm \\ \hline
        \textbf{DIRC-RAG Area} & 6.18mm$^2$ \\ \hline
        \textbf{Frequency} & 250MHz \\ \hline
        \textbf{Voltage} & 0.8V \\ \hline
        \textbf{Precisions} & INT4/8 \\ \hline
        \textbf{Embedding Dimension} & 128$\sim$1024 \\ \hline
        \textbf{Macro Size} & 16Kb \\ \hline
        \textbf{Macro Area} & 0.34mm$^2$ \\ \hline
        \textbf{Macro Efficiency} & 1176TOPS/W, 24.9TOPS/mm$^2$ \\ \hline
        \textbf{Macro NVM Storage} & 2Mb \\ \hline
        \textbf{Total NVM Storage} & 4MB \\ \hline
        \textbf{Total Memory Density} & 5.178Mb/mm$^2$ \\ \hline
        \textbf{Retrieval Latency} & 5.6$\mu$s (4MB retrieval) \\ \hline
        \textbf{Energy/Query} & 0.956$\mu$J (4MB retrieval) \\ \hline
    \end{tabular}
    }
    \vspace{-2mm}
    \label{tab:dirc_spec}
\end{table}

\begin{table*}[!ht]
    \centering
    \vspace{-2mm}
    \caption{Retrieval Precision (P@1, 3, 5) Comparison across Different Datasets and Quantizations.}
    \vspace{-2mm}
    \begin{threeparttable}
    \begin{tabular}{lcccccccccccc}
    \toprule
                 & \multicolumn{3}{c|}{Embedding Size (MB)} & \multicolumn{3}{c|}{P@1} & \multicolumn{3}{c|}{P@3} & \multicolumn{3}{c}{P@5} \\
    \cmidrule(lr){2-4} \cmidrule(lr){5-7} \cmidrule(lr){8-10} \cmidrule(lr){11-13}
    Dataset     & FP32      & INT8     & INT4   & FP32      & INT8     & INT4   & FP32      & INT8     & INT4   & FP32      & INT8     & INT4  \\ 
    \midrule
    SciFact     & 7.59      & 1.90     & 0.95   & 0.5067    & 0.5033   & 0.4833 & 0.2400    & 0.2378   & 0.2244 & 0.1633    & 0.1640   & 0.1553 \\ 
    NFCorpus    & 5.32      & 1.33     & 0.66   & 0.4210    & 0.4149   & 0.3684 & 0.3540    & 0.3488   & 0.3034 & 0.3046    & 0.3028   & 0.2743 \\ 
    TREC-COVID\textsuperscript{(1)}  & 15.68     & 3.92     & 1.96   & 0.6400    & 0.6200   & 0.5400 & 0.5667    & 0.5600   & 0.5533 & 0.5640    & 0.5520   & 0.4960 \\ 
    ArguAna     & 12.71     & 3.18     & 1.59   & 0.2525    & 0.2560   & 0.2489 & 0.1669    & 0.1650   & 0.1562 & 0.1255    & 0.1255   & 0.1172 \\ 
    SciDocs\textsuperscript{(2)}     & 12.53     & 3.13     & 1.57   & 0.2410    & 0.2400   & 0.2160 & 0.1907    & 0.1917   & 0.1683 & 0.1570    & 0.1572   & 0.1408 \\ 
    \bottomrule
    \end{tabular}
    \end{threeparttable}
    \begin{tablenotes}
        \footnotesize
        \item[1]     \textsuperscript{(1)} The TREC-COVID dataset is sampled by a factor of 16 for storing all INT8 quantized embeddings on DIRC-RAG.
        \item[2]     \textsuperscript{(2)} The SciDocs dataset is sampled by a factor of 3 for storing all INT8 quantized embeddings on DIRC-RAG.
    \end{tablenotes}
    \vspace{-3mm}
    \label{tab:precision}
\end{table*}

\begin{figure}[t]
    \centering
    \begin{subfigure}[t]{0.7\columnwidth} 
        \centering
        \includegraphics[width=\columnwidth]{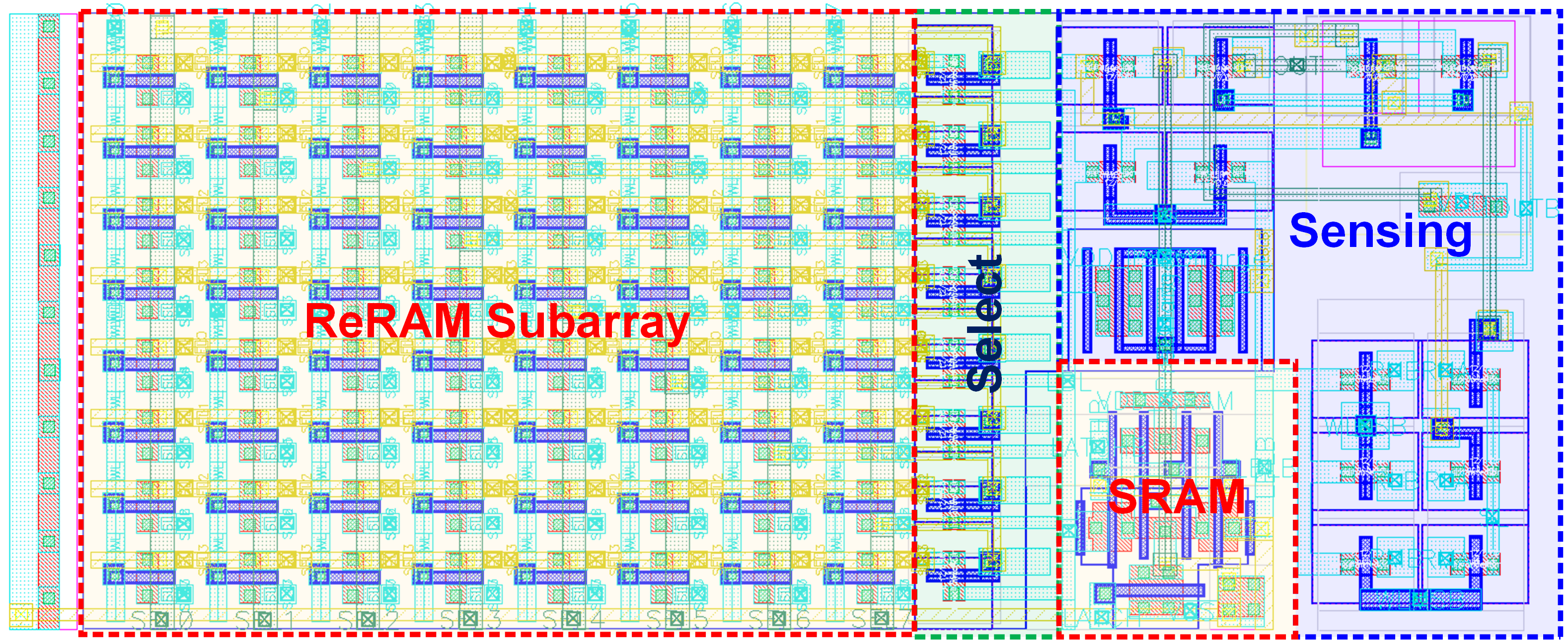} 
        \vspace{-4mm}
        \caption{DIRC cell layout}
    \end{subfigure}
    \begin{subfigure}[t]{0.7\columnwidth} 
        \centering
        \includegraphics[width=\columnwidth]{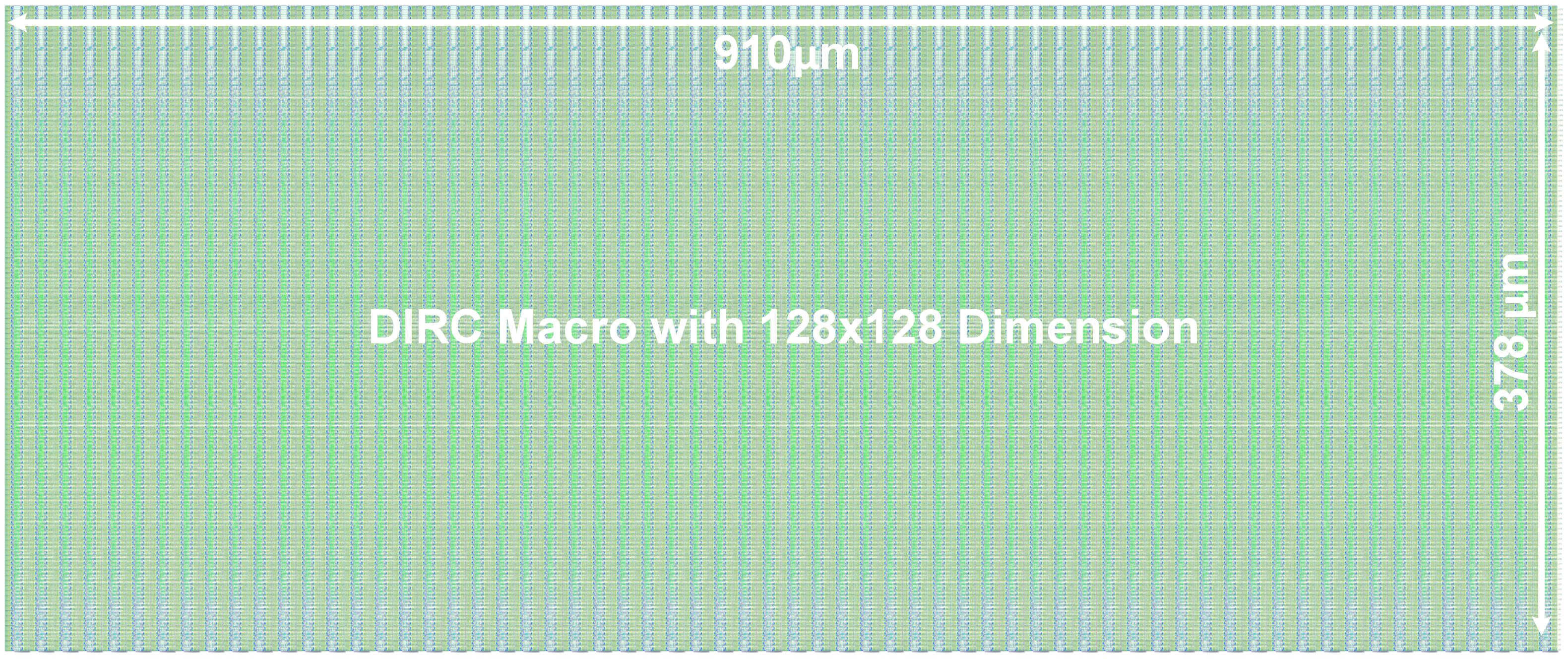} 
        \vspace{-4mm}
        \caption{DIRC macro layout}
    \end{subfigure}
    \caption{(a) The detailed layout of DIRC cell. (b) The layout overview of DIRC macro.}
    \label{fig7}
\end{figure}

For the software experiments, we modified the open-source \texttt{BEIR} framework \cite{beir2021} to evaluate the retrieval precision of INT4 and INT8 quantized embeddings \cite{ptq2018quantization} across datasets including SciFact \cite{scifact}, NFCorpus \cite{nfcorpus}, TREC-COVID \cite{treccovid}, ArguAna \cite{arguana}, and SciDocs \cite{scidocs}. The embedding model uses \texttt{all-MiniLM-L6-v2}, integrated into the open-source \texttt{Sentence-BERT} \cite{sentencebert2019} framework with an embedding dimension of 512. Retrieval performance is measured using Precision@k (P@k) for k = 1, 3, 5, reflecting the proportion of relevant documents in the top-k results. 

For system efficiency simulation, we developed a Python-based simulator to evaluate the energy efficiency and retrieval latency of the DIRC-RAG architecture. The simulator models energy consumption, interconnect bandwidth, and cycle latency for components such as the DIRC macro, norm unit, SRAM buffer, and global top-k comparator.

\subsection{Hardware Evaluation}
As shown in Figure 7, in this work, we designed a DIRC-RAG architecture capable of storing 4MB embeddings, supporting INT4/8 precision, and flexible dimensions ranging from 128 to 1024. We first conducted post-layout simulations for the DIRC macro. The results demonstrate that the DIRC macro, with an dimension of 128x128 and 2Mb of NVM storage, achieves an energy efficiency of 1176 TOPS/W and an area efficiency of 24.9 TOPS/mm\textsuperscript{2} at 250MHz and 0.8V.

We then performed system-level simulations for the overall DIRC-RAG with an area of 6.18 mm\textsuperscript{2}, which could accommodate 4MB of data, achieving a storage density of 5.178 Mb/mm\textsuperscript{2}. For a 4MB INT8-quantized document embedding database with 512 dimensions, our simulations reveal that querying the 4MB database requires only 5.6~$\mu$s of latency and 0.956~$\mu$J of energy consumption, while utilizing less than 1KB of SRAM buffer. Furthermore, we observed that the latency and energy consumption of DIRC-RAG scale linearly with the embedding database size.

However, it should be acknowledged that despite the high storage density achieved by DIRC-RAG, it is still insufficient. Two potential solutions could be adopted: First, when DIRC storage is insufficient, DIRC can function as a conventional SRAM-CIM for MAC operations by writing data from on-chip buffers or off-chip DRAM into SRAM; Second, scaling-up by leveraging chiplet technology to integrate multiple DIRC-RAG chips into a larger-scale system.

\subsection{Software Evaluation}

DIRC-RAG leverages a hardware-software codesign strategy by quantizing the query and document embeddings from FP32 to INT8 and INT4. To evaluate the impact of low-precision INT quantization on retrieval accuracy, we conducted extensive experiments on wide-range datasets and compared the results with the FP32 baseline. 

As shown in Table II, the retrieval precision of INT8 quantization is nearly identical to FP32, while INT4 introduces a slight but acceptable drop in P@k. For instance, on the NFCorpus dataset, P@1 decreases from 0.4210 (FP32) to 0.3684 (INT4), and P@5 from 0.3046 to 0.2743. With slight P@k drop, INT4 significantly boosts computational throughput and reduces storage compared to FP32. These results highlight that edge-based RAG systems can adopt low-precision INT quantization for document embeddings to simplify computation and minimize storage and bandwidth overhead while maintaining high retrieval accuracy.

\subsection{Comparison with Baseline}
Although \cite{qin2024robust} introduces CIM for RAG, it does not include a hardware evaluation. As there are few end-to-end edge RAG systems like DIRC-RAG available for a fair comparison of retrieval latency and efficiency, we conduct a performance evaluation against the NVIDIA RTX3090 GPU. As shown in Table III, for a single query in the SciFact dataset, DIRC-RAG with INT8 quantized embeddings achieves a latency of only 2.77~$\mu$s and an energy consumption of 0.46~$\mu$J. In contrast, the RTX3090 GPU, also using INT8 quantized embeddings, requires 21.7~ms of latency and 86.8~mJ of energy consumption. The GPU results are averaged over 30,000 queries to ensure accurate and reliable comparison.

\begin{table}[t]
    \centering
    \caption{Comparison with RTX3090}
    \vspace{-2mm}
    \resizebox{0.78\columnwidth}{!}{ 
    \begin{tabular}{|l|c|c|}
        \hline
        \textbf{Hardware} & \textbf{DIRC-RAG} & \textbf{RTX3090} \\ \hline
        \textbf{Process} & TSMC 40nm & Samsung 8nm \\ \hline
        \textbf{Area} & 6.18mm$^2$ & 628.4mm$^2$ \\ \hline
        \textbf{Frequency} & 250MHz & 1395MHz \\ \hline
        \textbf{Embeddings} & INT8 & FP32 \\ \hline
        \textbf{Dataset} & \multicolumn{2}{c|}{SciFact} \\ \hline
        \textbf{Precision@3} & 0.2378 & 0.2400 \\ \hline
        \textbf{Energy/Query} & 0.46 $\mu$J & 86.8 mJ \\ \hline
        \textbf{Latency/Query} & 2.77 $\mu$s &  21.7 ms \\ \hline
    \end{tabular}
    }
    \vspace{-4mm}
    \label{tab:dirc_vs_rtx3090}
\end{table}

\section{Conclusion}
In conclusion, DIRC-RAG effectively demonstrates a novel and efficient architecture for accelerating RAG on edge devices by leveraging digital in-ReRAM computation. Through the hybrid ReRAM-SRAM memory cell design and query-stationary dataflow, DIRC achieves high memory density, robust computational accuracy, and significant reductions in energy consumption and latency for retrieval. The proposed error optimization techniques further enhance the reliability of the system under device- and circuit-level variations. Simulation results validate the effectiveness of DIRC-RAG, achieving 5.6~$\mu$s/query latency, and 0.956~$\mu$J/query energy consumption, making it a promising solution for energy-efficient and high-performance RAG tasks on edge devices.

\clearpage

\bibliographystyle{unsrt}
\bibliography{reference}

\end{document}